\documentclass[runningheads,citeauthoryear]{apinv}
\usepackage{epsfig,cite,graphics}
\usepackage[T2A]{fontenc}
\usepackage[cp1251]{inputenc}

\begin{document}

\title{Light curve solutions of the eccentric \emph{Kepler} binaries KIC 11619964 and KIC 7118545 and mid-eclipse brightening of KIC 11619964
\thanks{based on data from the \emph{Kepler} mission}}
\titlerunning{Light curve solutions of KIC 11619964 and KIC 7118545}
\author{Diana Kjurkchieva\inst{1} and Doroteya Vasileva\inst{1}}
\authorrunning{D. Kjurkchieva, D. Vasileva}
\tocauthor{D. Kjurkchieva, D. Vasileva}
\institute{Department of Physics, University of Shumen, 115 Universitetska Str., Shumen, Bulgaria
	\newline
  \email{d.kyurkchieva@shu-bg.net}}
\papertype{Submitted on; Accepted on}
\maketitle

\begin{abstract}
We carried out light curve solutions of two eclipsing detached binaries on eccentric orbits observed by \emph{Kepler}. The orbits and fundamental parameters of KIC 11619964 and KIC 7118545 were determined with a high accuracy by modeling of their photometric data. We found that the temperatures of their components differ by around 2000 K while the radii of their secondaries are more than twice smaller than those of the primaries. We detected a strange "brightening" of KIC 11619964 in the narrow phase range ($\pm$ 0.0005) around the center of the primary eclipse reaching to 0.018 mag in amplitude. This "mid-eclipse brightening" needs follow-up observations with good time resolution.
\end{abstract}


\section*{Introduction}

The tidal forces change the stellar shape (tidal bulges) and cause brightness variability due to projection of the distorted stellar surfaces on the visible plane (Brown et al. 2011, Welsh et al. 2011, Morris 1985). It has double-wave shape (ellipsoidal variations) in the case of circular orbits and light increasing around the periastron in the case of eccentric orbits.

Close binaries on eccentric orbits are the main targets for study of the tidal phenomena: mechanisms for circularization of the orbits and synchronization of the stellar rotation with the orbital motion; impermanent mass transfer occurring close to the periastron (Sepinsky et al. 2007a, Lajoie \& Sills 2011); tidally excited brightening and oscillations (Kumar et al. 1995, Handler et al. 2002, Maceroni et al. 2009). The theoretical studies revealed that binaries could remain on eccentric orbits for long periods of time. Hence, the binary stars on eccentric orbits have also important evolutional role (Sepinsky et al. 2007b, 2009).

The eclipsing eccentric binaries (EEBs) with an apsidal motion provide valuable observational tests of the theoretical models of stellar structure and evolution (Kopal 1978, Claret \& Gimenez 1993, Willems \& Claret 2005). These stellar systems are important objects for the modern astrophysics but their study is straitened due to the long periods. Recently the huge surveys as ROTSE, MACHO, ASAS, SuperWASP, covering large part of the whole sky, increased considerably the number of EEBs. However, the huge contribution to their study belongs to the space missions, especially \emph{Kepler} (Koch et al. 2010), covering small sky area, but providing high-accuracy data. The unprecedented \emph{Kepler} observations allowed to discover and investigate a new tidally excited effect, called "heartbeat" phenomenon (Welsh et al. 2011, Thompson et al. 2012, Kjurkchieva \& Vasileva 2015).

Several thousands eclipsing detached systems, considerable part of them on eccentric orbits, were discovered by \emph{Kepler} (Prsa et al. 2011). The rich and valuable resources of the \emph{Kepler} database are available for additional research.

The goal of this study is determination of the orbits and physical parameters of two eccentric binaries, KIC 11619964 and KIC 7118545. They have relatively long eclipses (above 0.01 in phase units) and allow precise light curve solutions. Table 1 presents available information for these targets (Prsa et al. 2011, Slawson et al. 2011): orbital period $P$; \emph{Kepler} magnitude $m_{K}$; mean temperature $T_{m}$; width of the primary eclipse $w_{1}$ (in phase units); width of the secondary eclipse $w_{2}$ (in phase units); depth of the primary eclipse $d_{1}$ (in flux units); depth of the secondary eclipse $d_{2}$ (in flux units); the phases $\varphi_{2}$ of their secondary eclipses (the phases $\varphi_{1}$ of the primary eclipses are 0.0).

\begin{table}[htb!!!]
	\begin{center}
	\caption{Parameters of the targets from the EB catalog}
	\begin{tabular}{ccccccccc}
		\hline\hline
	\noalign{\smallskip}
Kepler ID &  $P$    & $m_{K}$ & $T_{m}$ & $w_{1}$ &  $w_{2}$ & $d_{1}$ &  $d_{2}$  & $\varphi_{2}$ \\
	\noalign{\smallskip}
	\hline
11619964 & 10.3685 & 14.545  & 5582   & 0.015 & 0.014   & 0.136  & 0.035   & 0.456   \\
7118545  & 14.7972 & 14.185  & 6095   & 0.019 & 0.014   & 0.251  & 0.029   & 0.675   \\
		\hline
	\end{tabular}
	\label{table1}
	\end{center}
\end{table}

\section*{Light curve solutions}

The modeling of the \emph{Kepler} data was carried out by the package \emph{PHOEBE} (Prsa \& Zwitter 2005). The out-of-eclipse parts of the observed light curves of the two targets are almost constant and we used for modeling the mode "Detached binaries".

We calculated preliminary values of the eccentricity $e$ and periastron angle $\omega$ by the formulae (Kjurkchieva \& Vasileva 2015)

\begin{equation}
e_0 \cos\omega_0=\frac{\pi}{2}[(\varphi_2-\varphi_1)-0.5]
\end{equation}

\begin{equation}
e_0 \sin\omega_0=\frac{w_2-w_1}{w_2+w_1},
\end{equation}

which are approximations of the formulae of Kopal (1978). The obtained values of $e_{0}$ and $\omega_{0}$ were used as input parameters of $PHOEBE$.

The mean temperatures $T_m$ of our targets (Table 1) required to adopt coefficients of gravity brightening 0.32 and reflection effect 0.5 (appropriate for stars with convective envelopes). We used linear limb-darkening law with limb-darkening coefficients corresponding to the stellar temperatures and \emph{Kepler} photometric system (Claret $\&$ Bloemen 2011).

We used for modeling 5000 points from the quarters Q1 and Q2 for each target and the period values from Table 1.

The procedure of the light curve solutions was carried out in several stages. Initially the primary temperature $T_{1}$ was fixed to be equal to the mean target temperature $T_{m}$ (Table 1). We input some guessed values of the secondary temperature $T_{2}$, mass ratio $q$, orbital inclination $i$ and potentials $\Omega_{1,2}$ (appropriate for detached systems) and varied only the eccentricity \emph{e} and
periastron angle $\omega$ around their input values $e_0$ and $\omega_0$ to search for the best fit of the phases of the eclipses (estimated by the value of $\chi^2$).

At the second stage we fixed \emph{e} and $\omega$ and varied simultaneously $T_{2}$, $q$, $i$ and $\Omega_{1,2}$ (and thus relative radii $r_{1,2}$) to search for the best fit of the whole light curves.

\begin{table}
\begin{center}
\caption{The derived orbital parameters of the targets}
\begin{tabular}{cccc}
\hline\hline
\noalign{\smallskip}
Kepler ID&   \emph{e }               & $\omega$ [deg]        & $\varphi_{per}$    \\
\noalign{\smallskip}
\hline
\noalign{\smallskip}
KIC 11619964 & 0.0891 $\pm$ 0.0001 & 223.45 $\pm$ 0.01 & 0.348      \\
KIC 7118545  & 0.3137 $\pm$ 0.0001 & 332.04 $\pm$ 0.01 & 0.769     \\
 \hline
\end{tabular}
\label{table2}
\end{center}
\end{table}

\begin{table}
\begin{center}
\caption{Parameters of the best light curve solutions}
\begin{tabular}{ccccccccc}
\hline\hline
\noalign{\smallskip}
Kepler ID& \emph{i}   & \emph{q  } & $T_1$  & $T_2$  &  $r_1$      & $r_2$      & $l_{1}$          & $l_{2}/l_{1}$  \\
\noalign{\smallskip}
\hline
\noalign{\smallskip}
11619964 & 88.306     & 0.602      & 5877     & 4177     & 0.0419      & 0.0187     & 0.950   & 0.0526  \\
         & $\pm$0.003 & $\pm$0.002 & $\pm$26  & $\pm$10  & $\pm$0.0002 & $\pm$0.0001&         &          \\
7118545  & 89.531     & 0.507      & 6154     & 3936     & 0.0408      & 0.0202     & 0.959   & 0.0427   \\
         & $\pm$0.001 & $\pm$0.001 & $\pm$12  & $\pm$3   & $\pm$0.0002 & $\pm$0.0003&         &         \\
 \hline
\end{tabular}
\label{table3}
\end{center}
\end{table}

Further, we used the obtained values $T_2$ (and correspondingly $\Delta T=T_m-T_2$) and $c=l_2/l_1$ ($l_2$ and $l_1$ are relative stellar luminosities from the second stage of the solution) to calculate the next approximations of $T_{1}^m$ and $T_{2}^m$

\begin{equation}
T_1^m=T_{\rm {m}} + \frac{c\Delta T}{c+1}
\end{equation}

\begin{equation}
T_2^m=T_1^m -\Delta T
\end{equation}

which are yet on the two sides of the mean value $T_m$ of the target.

Finally, we input the parameter values from the second stage and the new temperature values $T_{1}^m$ and $T_{2}^m$ and varied all parameters in small ranges around these values until reaching the best fit to the observations (minimum of $\chi^2$). The final parameters of the eccentric orbits are given in Table 2 while Table 3 contains the parameters of the stellar configurations. The synthetic curves corresponding to the parameters of our light curve solutions are shown in Figs. 1-2 as continuous lines.

The parameter errors in Tables 2--3 are the formal \emph{PHOEBE} errors. Their small values are natural consequence of the high precision of the \emph{Kepler} data.

The synthetic curves reproduced very well the \emph{Kepler} data. The residual curves show some bigger discrepancies during the eclipse phases (Figs. 1--2). Similar behavior could be seen also for other \emph{Kepler} binaries (Hambleton et al. 2013, Lehmann et al. 2013, Maceroni et al. 2014), especially those with small sum of relative radii. It was attributed to the effects of finite integration time (29.42 minutes for the \emph{Kepler} long-cadence data) studied by Kipping (2010).

\begin{figure}[htb!!!]
  \begin{center}
    \centering{\epsfig{file=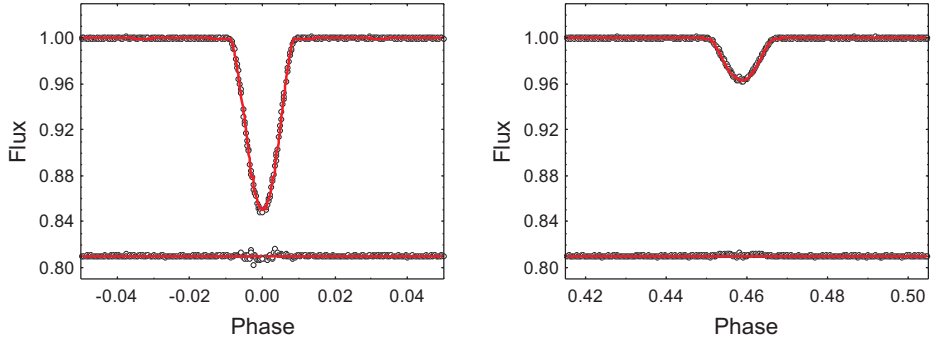, width=0.95\textwidth}}
    \caption[]{The primary (left panel) and secondary (right panel) eclipse of KIC 11619964 and their fits}
    \label{countryshape}
  \end{center}
\end{figure}

\begin{figure}[htb!!!]
  \begin{center}
    \centering{\epsfig{file=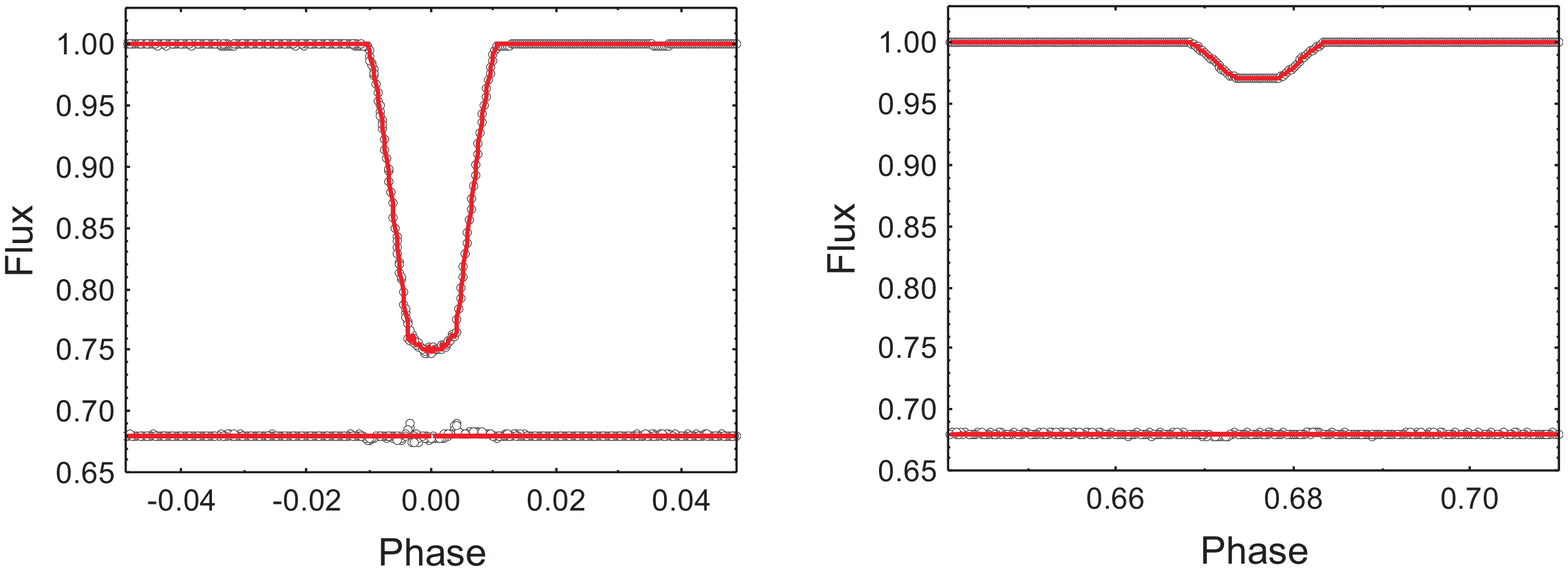, width=0.95\textwidth}}
    \caption[]{The primary (left panel) and secondary (right panel) eclipse of KIC 7118545 and their fits}
    \label{countryshape}
  \end{center}
\end{figure}

\section*{Analysis of the results}

The analysis of our light curve solutions of KIC 11619964 and KIC 7118545
led to several conclusions.

(1) The temperatures of stellar components are in the range
3930--6150 K. The primaries are the hotter components. The
temperatures of the secondaries are smaller with about 2000 K than
those of the primaries. This result is natural consequence of the
shallow secondary eclipses of the two targets.

(2) The orbital inclinations of the targets are near to
90$^\circ$ (Table 3) that is expected for eclipsing systems with
periods above 10 days. KIC 7118545 undergoes total eclipse.

(3) The stellar radii of the secondaries of the two targets are
more than twice smaller than those of the primaries. This result
together with their considerably lower temperatures lead to the
very small luminosity ratio $l_2/l_1$ for the two eccentric binaries
(0.043--0.053).

(4) The mass ratios of the targets are within the range 0.5--0.6.

(5) We did not find evidences for apsidal motion of our targets.
The possible reason is the relative short duration of the
\emph{Kepler} observations. Typically apsidal periods are at least
decade long (Michalska $\&$ Pigulski 2005). Moreover, the systems
with apsidal motions are with the shortest orbital periods or with
the largest sum of relative radii for a given eccentricity
(Michalska 2007) but these conditions are not fulfilled for our
targets.

(6) The review of the light curves of our targets from different
quarters did not exhibit any long-term variability.

(7) The out-of-eclipse light of the targets is constant within
0.15 $\%$.

\begin{table}
\begin{center}
\caption{Comparison of our results with those of automated fitting}
\begin{tabular}{ccccc}
\hline\hline
\noalign{\smallskip}
Kepler ID    & $T_2/T_1$   & $r_1+ r_2$ & $\sin i$ & source  \\
\noalign{\smallskip}
\hline
\noalign{\smallskip}
KIC 11619964 & 0.875  &  0.098   &  0.99486 &   Slawson et al. 2011\\
             & 0.711  &  0.0606  &  0.99956 &   our      \\
KIC 7118545  & 0.823  &  0.086   &  0.99855 &  Slawson et al. 2011\\
             & 0.640  &  0.0610  &  0.99997 &   our      \\
\hline
\end{tabular}
\label{table4}
\end{center}
\end{table}

(8) Table 4 presents the values of the temperature ratio $T_2/T_1$,
sum of the relative radii $r_1+r_2$ and $\sin i$ of our "manual"
light curve solutions and those determined by a neural network analysis (automated modeling)
of the phased light curves of our targets (Prsa et al. 2011, Slawson et al. 2011).
The last method does not provide error values but gives
statistical parameter uncertainties. Slawson et al. (2011)
estimated that 90$\%$ of the sample of detached and semi-detached
EBs had a corresponding error in $T_2/T_1$, $r_1+r_2$ and $\sin i$
smaller than 10 $\%$. However, our solutions do not confirm this
estimation, particularly for KIC 11619964 and KIC 7118545.

(9) We found a strange "brightening" (Fig. 3) of KIC 11619964
in the narrow phase range 0.0005 around the center of the primary eclipse
reaching to 0.018 mag in amplitude (these points were excluded from the procedure of modeling).
Unfortunately, there are not short-cadence data of KIC 11619964
(with good time resolution) and we are not able
to analyze the observed effect in details.

\begin{figure}[htb!!!]
  \begin{center}
    \centering{\epsfig{file=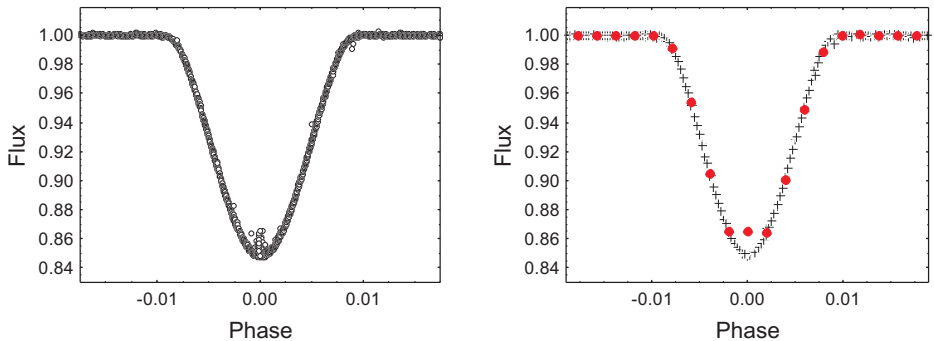, width=0.95\textwidth}}
    \caption[]{Left panel: The "brightening" at the primary eclipse of KIC 11619964 (all 65000 points available in the \emph{Kepler} archive are used); Right panel: primary eclipse with brightening (red circle symbols) superposed on several normal ones (black pluses)}
    \label{countryshape}
  \end{center}
\end{figure}

The detected mid-eclipse brightening could be attributed to: (i) artificial effect of the \emph{Kepler} observations or automated reducing or de-trending of the data of the \emph{Kepler} archive (but why only for this target?); (ii) some peculiarity of KIC 11619964. Similar effect has been established for the primary or secondary eclipse of other detached, semidetached,
contact and overcontact binaries (see table in Snyder $\&$ Lapham 2008).
Their mid-eclipse brightenings also do not occur at every eclipse (Pribulla 1999)
and have variable amplitude. The amplitudes of four of the detached systems
in the table of Snyder $\&$ Lapham (2008) are 0.002--0.04 mag, only that of $\varepsilon$ Aur
is considerably bigger. The phenomenon of mid-eclipse brightening
has not any plausible explanation yet (Snyder $\&$ Lapham 2008).

\section*{Estimation of the global parameters}

Due to the lack of radial velocity measurements we estimated the
global parameters of the target components by the following
procedure.

The primary luminosity $L_1$ was determined by the
relation "temperature, luminosity" for MS stars while the
secondary luminosity was calculated by the formula $L_2=(l_2/l_1)$$L_1$
where the luminosity ratio
$l_2/l_1$ is derived from the light curve solution (Table 3).

The orbital separation $a$ in solar radii was obtained from the
equation

\begin{equation}
\log a = 0.5 \log L_i - \log r_{i} - 2 \log T_{i} + 2 \log
T_{\odot},
\end{equation}

where the relative radii $r_{i}$ and temperatures $T_{i}$ were
taken from the light curve solution (Table 3). Then the absolute
radii were calculated by $R_{i}=ar_{i}$.

The total mass $M$ (in solar units) was calculated from the
third Kepler law

\begin{equation}
M=\frac{0.0134 a^3}{P^2}  ,
\end{equation}

where the orbital period \emph{P} was in days while the orbital
separation \emph{a} was in solar radii. Then the individual masses
$M_i$ were determined from the formulae $M_1= M/(1+q)$ and $M_2=M-M_1$.

\begin{table}
\begin{center}
\caption{Global parameters of the target (in solar units)}
\begin{tabular}{ccccccc}
\hline\hline
\noalign{\smallskip}
Kepler ID&  $M_1$         & $M_2 $         &  $R_1$          & $R_2$          &   $L_1$        & $L_2$       \\
\noalign{\smallskip}
\hline
\noalign{\smallskip}
11619964 &1.25 $\pm$ 0.09&0.75 $\pm$ 0.05  &1.05 $\pm$ 0.03  & 0.47 $\pm$ 0.01& 1.11 $\pm$ 0.03 & 0.058 $\pm$ 0.004  \\
7118545  &0.74 $\pm$ 0.03& 0.37 $\pm$ 0.01 & 1.07 $\pm$ 0.02 & 0.53 $\pm$ 0.01& 1.49 $\pm$ 0.01 & 0.064 $\pm$ 0.004  \\
 \hline
\end{tabular}
\label{table5}
\end{center}
\end{table}

\section*{Conclusions}

This paper presents the results of determination of the orbits and
fundamental parameters of the eclipsing eccentric binaries KIC 11619964
and KIC 7118545 on the basis of their \emph{Kepler} data.
The results could be used to improve the empirical
relations between the stellar parameters as well as to investigate
the tidal induced effects.

KIC 11619964 deserves follow-up photometric observations, especially
at the primary minimum, with good time resolution
to study its mid-eclipse brightening effect.

\section*{Acknowledgments}

The research was supported partly by funds of project RD-08-285 of
Scientific Foundation od Shumen University. It used the SIMBAD
database and NASA Astrophysics Data System Abstract Service. We
worked with the live version of the \emph{Kepler} EB catalog
(http://keplerebs.villanova.edu/). The authors are grateful
to the anonymous referee for the useful notes and propositions.

\end{document}